\documentclass[prc,aps,floatfix,groupedaddress,amsmath,amssymb]{revtex4}
\usepackage{graphicx}
\usepackage{epsfig}

\def\beq{\begin{equation}}
\def\eeq{\end{equation}}
\newcommand{\rv}{\mathbf{r}}

\newcommand{\be}{\begin{eqnarray}}
\newcommand{\ee}{\end{eqnarray}}

\newcommand{\een}{\nonumber\end{eqnarray}}

\begin{document}
\thispagestyle{empty}
\vspace*{0.5 cm}
\begin{center}
\title{Beyond the mean field in the particle-vibration coupling scheme.}
\author{\bf M. Baldo$^1$, P.F. Bortignon$^{2,3}$, G. Col\`{o} $^{2,3}$, D. Rizzo $^2$ and L. Sciacchitano$^2$
\footnote{Present address : Pharma Quality Europe (PQE), via B. Marcello 36, I-20133 Milano, Italy}}  
\affiliation{$^1$ INFN, Sezione di Catania, via S. Sofia 64, I-95123, Catania, Italy}
%
\affiliation{$^2$ Dipartimento di Fisica, Universit\`a degli Studi di Milano, via Celoria 16, I-20133 Milano, Italy}
\affiliation{$^3$ INFN, sezione di Milano, via Celoria 16, via Celoria 16, I-20133 Milano, Italy}
%
\maketitle

\vspace*{1 cm}
\end{center}
{\bf ABSTRACT} \\
\noindent The Energy Density Functional theory is one of the most used methods developed in nuclear structure. It is based on the assumption that the energy of the ground state is a functional only of the density profile. The method is extremely successful within the effective force approach, noticeably the Skyrme or Gogny forces, in reproducing the nuclear 
binding energies and other bulk properties
along the whole mass table. Although the Density Functional is in this case represented formally as the Hartree-Fock mean field of an effective force, the corresponding single particle states in general do not reproduce the phenomenology particularly well. To overcome this difficulty, a strategy has been developed where the effective force is adjusted to reproduce directly the single particle energies, trying to keep the ground state energy sufficiently well reproduced. An alternative route, that has been developed along several years, for solving this problem is to introduce the mean field fluctuations, as represented by the collective vibrations of the nuclear system, and their influence on the single particle dynamics and structure. This is the basis of the particle-vibration coupling model. 
In this paper we present a formal theory of the particle-vibration coupling model based on the Green' s function method. The theory extends to realistic effective forces the macroscopic particle-vibration coupling models and the (microscopic) Nuclear Field Theory. It is formalized within the functional derivative approach to many-body theory. 
An expansion in diagrams is devised for the single particle self-energy and the phonon propagator. Critical aspects of the particle-vibration coupling model are analysed in general. Applications at the lowest order of the expansion are presented and discussed.  

\maketitle

\section{Introduction}
One of the most striking features of nuclei is the validity of the shell model where the nucleons move
independently in a smooth single-particle (s.p.) potential.This potential is static, but non-local in space, which
implies that the mean field is velocity (momentum) dependent. The fluctuations of this average potential leads to
collective modes,in particular surface vibrations (phonons) \cite{BM}. Taking into account the coupling of these modes to the single
particle motion, as in the particle-vibration coupling (PVC) model, the shell model acquires a dynamical content. Thus, the average potential becomes also non
local in time, being characterized by an energy (frequency) dependence. The dynamics of the shell model (the
expression "dynamics of the shell model" is after G.E. Brown \cite{Gerry}) affects different nuclear properties as
fragmentation and related spectroscopic factors of s.p. states, their density (proportional to the effective mass m*
near the Fermi energy), the s.p. spreading widths, the imaginary component of the optical potential. A unified
description of the s.p. motion at positive and negative energy eventually emerges \cite{CM}.
The importance of the dynamical coupling was first demonstrated in the case of electrons in metals and of normal
liquid $^3$He. These results inspired the work of Bertsch and Kuo \cite{BK} on the enhancement of the effective mass near
the Fermi energy in finite nuclei. The review article in ref. \cite{CM} contains the many results obtained in the '80.
A recent introduction to the subject can be found in ref. \cite{PFB}. Extensions of the PVC approach to the continuum \cite{cont} and to the scattering processes \cite{scat} have been presented in the literature.  
\par
Most of the calculations done in the past are based on purely phenomenological inputs. In Tables 4.3a and 4.3b of
Ref. \cite{CM}, an extensive review of the results obtained in $^{208}$Pb by nine groups in the period 1968-1983 can be
found. Rather different frameworks had been adopted, no one being fully self-consistent: s.p. potentials range
from harmonic oscillator (HO) to Woods-Saxon (WS) or Hartree-Fock (HF) with Skyrme forces; residual interactions at the
particle-vibration vertex are either multipole-multipole forces, or forces of Landau-Migdal type, or Skyrme forces
but with velocity-dependent terms dropped, or even G-matrix interactions. Consequently, although there is
qualitative agreement among several calculations, it is rather hard to assess seriously the quantitative impact of
all the approximations done and compare in detail with EDF-based calculations with so called   "spectroscopic"
accuracy and with the recent relativistic calculations \cite{Ring,Afan1,Afan2}. These last one are fully self-consistent , because
the whole interaction is considered both to build the phonons  within the relativistic random phase approximation (RRPA)
framework and to construct the PVC vertex.
Within the framework of non-relativistic effective interactions, a step in this direction has been   done very
recently in ref. \cite{Colo2010,inpress} with calculations of $^{40}$Ca and $^{208}$Pb s.p. self-energies  based on a fully
self-consistent random phase approximation (RPA)  of the vibrations, and without dropping any term in the coupling vertex.

In this paper, along the scheme presented in ref. \cite{LNC}, we derive coupled integral equations for the single particle self-energy and Green' s function, the phonon propagator and the
vertex function. These equations form a closed system and
we propose a solution by iteration. The terms so obtained by iteration can be represented by diagrams,
that automatically contain the correct symmetry factors, as well as the Pauli corrections demanded by antisymmetry between the explicit particle lines and the ones implicitly contained in the phonon propagator. 
We propose a solution for the single-particle self-energy at the lowest order of approximation, 
with numerical applications 
for both density-independent and density-dependent forces. \par 
The outline of the paper is as follows. In Section \ref{sec:form} we 
present the basic formalism, starting from the construction of the 
hamiltonian in second quantization, from which we derive the equation 
of motion for the single particle Green's function and the integral 
equations for the phonon propagator and the vertex function. For 
simplicity in this section the treatment is restricted to density 
independent forces. In Section \ref{sec:exp} the expansion in diagrams 
is devised following an iterative procedure. Here some peculiarities 
of the formalism, which are consequences of the microscopic structure 
of the phonons, are discussed in some detail. Section \ref{sec:ddep} 
is devoted to the case of a density dependent force. We show that, 
 at least for the applications at lowest order that are discussed in this 
work, 
this case can be reduced to the density independent case if the phonons 
are assumed to be of small amplitude. In Section \ref{sec:app} we present the application of the formalism to the nucleus $^{40}$Ca, and compare the results for a density independent and a density dependent Skyrme force.
Conclusions are drawn in Section \ref{sec:conc}.

\section{The formalism\label{sec:form}}
In the general Energy Density Functional (EDF) approach it is assumed that the energy of the ground state can be obtained by the minimization of a functional that depends only on the single particle density profile. The method is based on the Hohenberg-Kohn theorem \cite{HK}, that proofs the existence of such a functional, although it does not provide any procedure to construct it. The theorem has been
recently extended \cite{self1,self2,self3} to self-bound systems like nuclei. In nuclear physics the most widely used scheme to formulate the functional is based on the introduction of effective forces,
noticeably  Skyrme or Gogny forces \cite{Bender}. These forces are phenomenological and are assumed to be used in the mean field scheme and to incorporate part of the effect of correlations through phenomenological parameters to be fitted to reproduce the binding energy of nuclei, as well as the trend of radii and deformations, throughout the mass table. 
Other functionals, e.g. the  recent BCPM \cite{BCP1,BCP2,BCP3}, contains more microscopic input,
but they still include phenomenological terms to be fitted. Not only masses, but also other
simple functionals of the density like radii etc. can be well
reproduced. These functionals have also shortcomings, and in
this work we focus on the fact that observables like the single-particle
states and their fragmentation cannot be accounted for.
The approach has been extremely successful and an enormous literature exists on the subject. Although in what follows we do not need the particular form of the functional,     
let us consider the case of a functional that can be formally written 
as the Hartree-Fock mean value of an effective force $f$,
\beq
E \,=\, T \,+\,  \frac{1}{2} \sum_{iklj} \rho_{ik} f_{il ; kj}^A(\rho) 
\rho_{lj} \,=\, T \,+\, V(\rho), 
\label{eq:fun}\eeq
\noindent where $T$ is the kinetic energy, $\rho_{ik}$ are the matrix elements of the density,  and $i j k .....$ are labels
for generic single particle states. The effective force $f$ is in general a functional of the density matrix $\rho_{ik}$. The matrix elements  of the effective force are antisymmetrized 
%
\beq
f_{il ; kj}^A  \,=\, <\, i l\, |\, f\, |\, k j\, >  
\,-\,  <\, i l\, |\, f\, |\, j k\, >.  
\label{eq:antif}\eeq
The density matrix can be written as the expectation value
\beq
\rho_{ik} \,=\, < \psi_i\,^\dag \psi_k >
\label{eq:ro}\eeq
\noindent where $\psi^\dag$, $\psi$ are the single particle creation and annihilation operators. 
The density dependence can originate from three-body (or higher) forces, if one replaces a pair
$\psi_i^\dag \psi_k$ by its ground state average, Eq. (\ref{eq:ro}), thus reducing the three-body force to density dependent two-body force. However in general the force contains non-integer power of the density, and it cannot be derived from a many-body force. Notice that the antisymmetrization in 
Eq. (\ref{eq:antif})is not performed with respect to the orbitals of the density matrix that appears implicitly in the effective force. This is a general problem for the effective density dependent force, discussed by several authors \cite{Stri,Cham}, but we follow the common procedure to use expression (\ref{eq:fun}). We will develop the formalism assuming a density independent force like SV \cite{SV}.
Later we will discuss how to implement the approach to the case of a density dependent force and we will discuss the problems that arise in that case. There are currently several attempts to fit functionals that are density independent, with the goal of avoiding problems that have been identified in the multi-reference DFT \cite{nonl}. 

\par The dynamics in the single particle degree of freedom can be described by the one-body Green' s
function and the corresponding equations of motion. To this purpose one needs a Hamiltonian formulation. We will
assume that the independent single particle motion is determined by the single particle potential, 
 while the coupling with the
density fluctuations is determined by the same interaction appearing in the functional 
of Eq. (\ref{eq:fun}), with $f$ independent of density. In other words the Hamiltonian will contain a one-body term,
which includes the single particle potential, and a residual two-body interaction which describes the fluctuation dynamics around the ground state. 
Therefore we will assume the following form of the Hamiltonian
\beq
 H  \,=\, H_0 \,+\, \frac{1}{4} \sum_{klmn} <\,k\,\, l\, |\,\, f\,\, | \, m\,\, n\, >_A
 N\big(\psi_k^\dag \psi_l^\dag \psi_n \psi_m \big)
\label{eq:ham}\eeq
\noindent
 The second quantization form of the Hamiltonian 
is used in order to incorporate in a systematic way the Pauli
principle in the equation of motion.
The symbol $N(....)$ indicates the normal product of the operators with respect to the independent particle ground state. This form
ensures that the two-body interaction has vanishing mean value in this ground state.
\noindent  
The term $H_0$ is the one-body part of the Hamiltonian, that in general can be written
\beq
 H_0 \,=\, \sum_{ik} \epsilon _{ik} \psi_i^\dag \psi_k
\label{eq:one}\eeq
\noindent  Here $\epsilon_{ik}$ is the single particle energy matrix, obtained from the Hartree-Fock approximation. This one-body part describes the single particle motion in the independent particle limit, i.e. when the two-body interaction term is neglected. 
The form of Eq. (\ref{eq:ham}) for the hamiltonian in the equation of motions poses however serious problems in general, and in particular if one wants to extend the formalism beyond RPA. 
First of all any functional is supposed to include at least the effect of the short range correlations in the ground state, as produced by the hard core of the bare NN interaction. In a microscopic approach this can be achieved by considering a two-body G-matrix in a restricted model space 
as the effective force. For a more phenomenological functional one can include any sort of correlation implicitly and have double counting problems. 
This possible double-counting can be bypassed in a purely phenomenological approach, where the force is re-fitted to reproduce the ground state properties and the single particle observables after the particle-vibration coupling is introduced at different level of approximation \cite{Doba_PVC}.   
Notice that in the particle-vibration coupling model the particle-particle and hole-hole scattering processes are not considered, since the dominant correlation is supposed to be determined by the density fluctuations. 
\par 
In the Hartree-Fock approximation the occupation number $n_i$ equals $1$ for occupied states (holes) and $0$ for non occupied states (particles). 
A partial justification of using the HF energies for the single particle orbitals, where the nucleons can move, stems from the Koopmans' theorem \cite{Koop}, valid for the HF approximation, which states that in linear approximation the energy for removing or adding a particle coincides indeed with the HF energy calculated in the core. However,
adding or removing a particle produces a variation on the Center of Mass corrections (CMC), since they depend on the mass number. The theorem is then violated and the CMC must be properly treated \cite{Doba}, but in this presentation we are not going to treat this problem. Other physical effects
can be relevant for the correct description of the single particle potential. If one particle is added or removed, the one-body potential and the single particle energies are expected to be modified ("core polarization").  
Our strategy is to incorporate these physical effects on the same footing as the dynamical effects on the single particle motion due to the particle-vibration coupling. This will be apparent along the development of the formalism.       
\par
We will use the Hamiltonian of Eq. (\ref{eq:ham}) to derive the equation of motion of
the single particle Green 's function and its coupling with the phonons of the nuclear system. 
To simplify the presentation and the formalism we will consider a generic single particle basis.
The normal product can be expanded as follows
\beq
\begin{array}{ll}
   N\big(\psi_k^\dag \psi_l^\dag \psi_n \psi_m \big) \,=\, \psi_k^\dag \psi_l^\dag \psi_n \psi_m
   &\,+\, <\psi_k^\dag\psi_n>_0 \psi_l^\dag\psi_m  \,-\, <\psi_k^\dag\psi_m>_0 \psi_l^\dag\psi_n \\
   \ &\, \\
   &\,-\, <\psi_l^\dag\psi_n>_0 \psi_k^\dag\psi_m  \,+\, <\psi_l^\dag\psi_m>_0 \psi_k^\dag\psi_n \\
   \ &\, \\
   &\,+\, <\psi_k^\dag\psi_m>_0 <\psi_l^\dag\psi_n>_0  \,-\, <\psi_k^\dag\psi_n>_0 <\psi_l^\dag\psi_m>_0 \\
\end{array} \label{eq:np}\eeq
\noindent
The subscript $0$ indicates mean value in the independent particle ground state. It is easy to verify that in the single particle basis that diagonalizes the one-body hamiltonian the ground state mean value of the normal product vanishes. It will vanish for a generic single particle basis, provided the mean values are still considered in the independent particle ground state. 
As we will see, the formal introduction of the normal product has a relevant consequence
in the development of the formalism. In fact
the single particle self-energy, appearing  in
the Green' s function beyond the HF approximation,
introduces modification of the single particle
density matrix and of the static mean field
acting on each particle. These modifications
are embodied in a definite term in the equations
of motion, see Eqs. (\ref{eq:self},\ref{eq:Upot}) below. In this
way one achieves a clear separation between 
the static polarization of the core, which
affects the single particle energy and
wave function \cite{Tarpa}, and the dynamics
of the nuclear system in the particle-vibration
coupling scheme.

\par
We introduce the single particle Green' s function $G_{pq}(t_1,t_2)$ according to the usual definition
\beq
 G_{pq}(t_1,t_2) \,=\, -i <T\{\psi_p(t_1) \psi_q^\dag(t_2)\}>
\label{eq:green}\eeq
\noindent where the creation and annihilation operators are in the Heisenberg representation and the mean value is
in the correlated ground state. The equation of motion for $G$ can be derived following the standard procedure. 
As shown in
the Appendix the equation can be derived using the functional derivative method \cite{BaymKad,Baym}. In this
scheme one introduces a single particle external potential $\phi(t)$, that depends on the time $t$, and the
dynamical evolution of the system under this external potential is analysed. The functional derivatives of
different many-body quantities, in particular the single particle Green' s function, introduce systematically the
correlation functions that characterize the dynamics of the system. In particular one has
\beq
 -i <T\{\psi_k^\dag(t_1)\psi_m(t_1)\psi_n(t_1)\psi_q^\dag(t_2)\}> \,\,=\,\,
 i \frac{\delta G_{nq}(t_1,t_2) }{ \delta \phi_{km}(t_1)}
 \,+\, <\psi_k^\dag(t_1)\psi_m(t_1)> G_{nq} (t_1,t_2)
\label{eq:exder}\eeq
 When all the functional derivatives are
calculated at $\phi(t) \,=\, 0$, the equation of motion of the correlation functions of the many-body system, in
absence of perturbation, are readily obtained. In this way one gets a set of equations that couple the
correlation functions that is closed, and it can be the starting point of different schemes of
approximation.
 The method
was previously used in refs. \cite{LNC,HedLun,Hed}.  
In this
paper 
we proceed further in the development of the formalism and we will discuss in the next sections its extension to the case of a density dependent force. 
Details of the derivation can be found in the Appendix, while here we describe the main features of the formalism and the resulting basic equations. The equation of motion of the single particle Green' s function can be written in the form (Dyson' s equation) 
\beq
i \frac{\partial }{ \partial t_1} G_{pq}(t_1,t_2) \,-\, \sum_{p'}\epsilon_{pp'} G_{p'q}(t_1,t_2) \,-\, \sum_h
M_{ph}(t_1,\overline{t_3}) G_{hq}(\overline{t_3},t_2) \,=\, \delta(t_1 - t_2)\delta_{pq}
\label{eq:Dyson}\eeq
\noindent 
where $\epsilon_{pp'}$ is the 
the single particle energy matrix, previously introduced, and all the higher order contributions have been collected  in the self-energy $M_{ph}$. In this equation, and in the following,
a line over a time variable indicates integration over that variable (following the notation of
ref. \cite{BaymKad}). 
The self-energy $M_{ph}$ can be further separated in the sum of two terms
\beq
M_{ph}(t_1,t_3) \,=\, U_{ph}(t_1) \delta(t_1 - t_3) \,+\, \tilde{M}_{ph}(t_1,t_3)
\label{eq:self}\eeq
The first one is the polarization potential $U$ \cite{LNC,HedLun,Hed}
\beq
 U_{ph}(t_1) \,=\, \phi_{ph}(t_1)
 \,+\, \sum_{km}   <\, k\,\, p\, |\,\, f\,\, |\, m\,\, h\,>_A \big[ <\psi_k^\dag(t_1) \psi_m(t_1)
> \,-\, <\psi_k^\dag(t_1) \psi_m(t_1)>_0  \big]
\label{eq:Upot}\eeq
\noindent
where the brackets $< ... >_0$ indicate mean value in the ground state of $H_0$, while $< ... >$ indicate mean value in the correlated ground state. This polarization potential takes into account the effect of the fluctuations of the density matrix with respect to the HF one.
Therefore the residual interaction contributes to the static single particle potential only as a correction to the original Hartree-Fock single particle potential. This is a direct consequence of the normal product form of the Hamiltonian of Eq. (\ref{eq:ham}), which ensures that the residual interaction has vanishing mean value in the unperturbed ground state.      
It is important to remark that the polarization potential is non zero whenever the calculation is beyond the original approximation for the single particle energies that appear in the free Hamiltonian $H_0$ of Eq. (\ref{eq:one}). 
 Indeed it involves the difference between the higher
order density matrix and the one corresponding to the one-body part of the Hamiltonian. The correlated density matrix can be evaluated, in principle, for any given
approximation for the Green' s function. This contribution is just due to the correction to the original static single particle potential produced by
the correlations in the ground state, as previously anticipated, as well as the rearrangement of the single particle potential if one particle is added (or removed) to the nucleus (the static "core polarization"). 
\par
The second term in Eq. (\ref{eq:self}) describes the dynamical contributions not included in the static single particle potential. It can be written
\beq
\tilde{M}_{ph}(t_1,t_3) \,=\, M_{ph}'(t_1,t_3) \,+\, \frac{1}{2} \sum_{km} <\, k\,\, p\, |\,\, f\,\, |\, h\,\, m\,>_A
 < \psi_k^\dag(t_1) \psi_m(t_1)>\delta(t_1 - t_3)
\label{eq:Mtil}\eeq
\noindent where
\beq
M_{ph}'(t_1,t_3) \,=\, \frac{1}{2} i \sum_{nrij} W_{pn ; ij}(t_1,\overline{t_5}) G_{nr}(t_1,\overline{t_4}) \Gamma_{rh ;
ij}(\overline{t_4},t_3 ; \overline{t_5})
\label{eq:Mpri}\eeq
The self-energy part $M'$ is the one that includes the dynamical polarization processes (dynamical "core polarization"). In
fact the quantity $W$ can be written
\beq
W_{pn ; ij}(t_1,t_5) \,=\, <\, i\,\, p\, |\,\, f\,\, |\, j\,\, n\,>_A \delta(t_1 \,-\, t_5)
 \,-\, i\sum_{kmrs} <\, k\,\, p\, |\,\, f\,\, |\, m\,\, n\,>_A \Pi_{mk ; rs}(t_1,t_5)
 <\, r\,\, i\, |\,\, f\,\, |\, s\,\, j\,>_A
\label{eq:W}\eeq
\noindent and $\Pi$ is the particle-hole response function of the nuclear system
\beq\begin{array}{ll}
\Pi_{sr ; km}(t,t') &\, =\, -\, <\,\Psi_0\, | T\{\rho'_{rs}(t) \rho'_{km}(t')\} | \,\Psi_0\, >\, =\,
 -\,  <\,\Psi_0\, | T\{\rho'_{sr}(t)^\dag \, \rho'_{km}(t')\} | \,\Psi_0\, > \\
\ &\,  \\
\ \ \ \ \ \ \ \ \rho'_{rs}(t) &\,=\, \psi_r^\dag (t) \psi_s (t)
\, -\, < \psi_r^\dag \psi_s  > \\
\end{array}\label{eq:pol}\eeq
\noindent It is through the response function $\Pi$ that the phonon degree of freedom is introduced in the
formalism. Its coupling with the particle degree of freedom (i.e. the Green' s function) that appears in
Eq. (\ref{eq:Mpri}) describes the elementary process in which the fluctuations of the single particle potential affects the particle propagation.  The vertex function $\Gamma$ takes into account the effects not included in this coupling process of
the particle with the phonon. These effects are usually indicated as "vertex
corrections". However, the present formalism describes these effects still in terms of more complex processes of particle-vibration coupling. As shown in the Appendix, the phonon propagator $\Pi$ and the vertex function $\Gamma$ satisfy the following two relations
that close the set of equations for $M$, $\Pi$ and $\Gamma$
\beq\begin{array}{ll}
\ \ \ \ \Pi_{mk , rs}(t,t') &=\,\, \sum_{pqij} G_{mp}(t,\overline{t_1}) G_{qk}(\overline{t_2},t)\Gamma_{pq ; ij}
(\overline{t_1},\overline{t_2} ; \overline{t_3})\times \\
\ &\,  \\
&\big[\, \delta_{ir}\delta_{js}\delta(\overline{t_3} - t') - i \sum_{tu}  <\, i\,\, t\, |\,\, f\,\, |\, j\,\,
u\,>_A \Pi_{ut , rs}(\overline{t_3},t')\, \big]  \\ 
\ &\, \\
\ &\, \\
\Gamma_{nh ; ij}(t_3,t_4 ; t_5) &=\,\, \delta_{ni}\delta_{hj}\delta(t_3 - t_4)\delta(t_3 - t_5) \,-\, \sum_{rspq} \big( \frac{\delta
\tilde{M}_{nh}(t_3,t_4) }{ \delta G_{rs}(\overline{t_6},\overline{t_7}) } \big)
G_{rp}(\overline{t_6},\overline{t_8})
G_{qs}(\overline{t_9},\overline{t_7}) \Gamma_{pq ; ij}(\overline{t_8},\overline{t_9} ; t_5) \\ 
\end{array}\label{eq:pigam}\eeq
\par 
\noindent Together with the Dyson Eq. (\ref{eq:Dyson}) and the expression (\ref{eq:W}) for $W$, they form a closed set of coupled equations.
Equations (\ref{eq:Mpri},\ref{eq:W},\ref{eq:pol},\ref{eq:pigam}) has been introduced by Hedin
\cite{HedLun,Hed}, see ref. \cite{Onida} for a more recent presentation. In the realm of nuclear
physics this scheme has been considered in ref. \cite{Frac}. 
 We stress that the vertex function $ \Gamma $ is irreducible, i.e. it  does not contain  any polarization propagator starting at the vertex of time 
$ t_5 $. In fact this would produce an over-counting of contributions in the equation (\ref{eq:pigam}) for the phonon propagator
and unphysical higher order poles \cite{Kamer}, which are in fact not present in the expansions of these equations proposed in the literature, see e.g. ref. \cite{Onida,Godby}. As already mentioned, it is in this sense that, in in the terminology of Ref. \cite{HedLun,Hed,Onida}, the vertex function $ \Gamma $ includes only the ''vertex corrections".\par
\par
In the integral equation for $\Gamma$ the kernel involving the functional derivative of $\tilde{M}$ with
respect to $G$ can be considered as the effective interaction that determines the vertex corrections.  Solving these coupled equations is equivalent to the requirement of the self-consistency
among $M$, $\Pi$ and $\Gamma$. In particular one should solve the Dyson equation for $G$. It is important to notice that if the equations are solved in an approximation that goes beyond the independent particle approximation, implicitly some correlations are introduced in the ground state.
In general, in the Green' s function formalism the explicit ground state wave function is not introduced, but the one-body (from the Green' s function) and the two-body (from the phonon propagator) density matrices can be extracted. As mentioned before, these correlations modify also the single particle static potential through the polarization potential $U$ of Eq. (\ref{eq:Upot}). The single particle potential will not correspond any more to the one-body part of the Hamiltonian. 
It is still possible to get the energy of the correlated ground state  
by means of the Koltun' s sum rule \cite{RS}, since it involves only the single particle Green' s function. However this is not within the scope of the present paper. \par 
We close this section by mentioning the possible extension of the method to include the pairing correlations.  Besides the external potential $ \phi $ coupled to the density, one can in fact introduce an external source coupled to the pairing fields $ \psi_i \psi_j $ and $ \psi_i\,^\dag \psi_j\,^\dag $.
Starting from the anomalous Green' s functions,  
$ -i <T\{\psi_p(t_1) \psi_q(t_2)\}> $ and $ -i <T\{\psi_p^\dag(t_1) \psi_q^\dag(t_2)\}> $,
one can generate the propagators and vertex functions pertinent to the pairing correlations, and the corresponding extended set of coupled equations. This extended formalism will be discussed elsewhere, and we restrict here to the normal systems.
\par
\vskip 0.4 cm
\section{Expansion in diagrams and their estimate\label{sec:exp}}
\par
Once the general formalism is established, one can formulate different approximation schemes. Generally speaking one can solve this set of equations by iteration, starting from a guess on the first order approximation. One can start from an approximate expression for $\Gamma$ and substitute it in Eq. (\ref{eq:Mpri}) and (\ref{eq:pigam}), which provides an approximation for $\tilde{M}$ and $\Pi$. The functional derivative of this self-energy is then used to get a new approximation for $\Gamma$. Alternatively, one can start with an approximation for $\tilde{M}$, perform the functional derivative and continue the procedure.  
One could wonder about the second term in the expression  (\ref{eq:Mtil}) for $\tilde{M}$, which  is a mean value not yet specified in the first step of the iteration. However, we will see explicitly in the following that this term is cancelled by an identical term arising in the expansion of $\tilde{M}$.   
This cancellation is exact, and no ambiguity is present in the formalism.     
\par     
The iteration provides an expansion of the basic quantities $G$, $\Pi$ and $\Gamma$, and each term of the expansion can be represented by diagrams. At each step of the iteration the so-obtained approximation for $\Gamma$, $\tilde{M}$ and $\Pi$ is conserving, i.e. the local current is conserved. As shown in ref. \cite{BaymKad}, this property follows from 
the fact that the effective interaction in 
Eq. (\ref{eq:pigam}) for $\Gamma$ is obtained by functional derivative of $\tilde{M}$. Notice that the Eq. (\ref{eq:pigam}) for $\Pi$ is also modified at each step of the iteration. In principle the starting point is arbitrary, since all possible terms of the expansion are automatically generated in any case.    
\par 
As suggested by the above discussion, the simplest approximation is to take $\tilde{M} \,=\, 0$ and $\Gamma \,=\, 1$. In this way one gets the Random Phase Approximation (RPA) for the phonon propagator $\Pi$, with the single particle Green's functions calculated in the independent particle approximation. This is known to be a conserving approximation. \par   
In this approximation the set of equations is solved in a consistent way.
In fact, in our scheme the single particle energies of the one-body Hamiltonian are fixed and their functional derivative vanishes. 
 The RPA describes with good accuracy the low lying and high lying collective vibrations. However the corresponding widths require the introduction of more complex configurations beyond the standard RPA \cite{RMP}, see later.  
 One can then continue the iteration procedure by inserting $\Gamma = 1$ in the general expression (\ref{eq:Mpri})
for the self-energy, perform the functional derivative to be inserted in the equation for $\Gamma$, and so on.  
\par 
Of course a standard perturbation expansion in terms of the interaction $f$ could be also performed, but the scheme of the described iterative procedure
involves both the correlated single particle Green' s function and phonon propagator (at the given order of the expansion) and at each level of the iteration the corresponding approximation is conserving.
 \par
\begin{figure}[b]
%
\begin{center}
\includegraphics[bb= 240 0 400 790,angle=0,scale=1.]{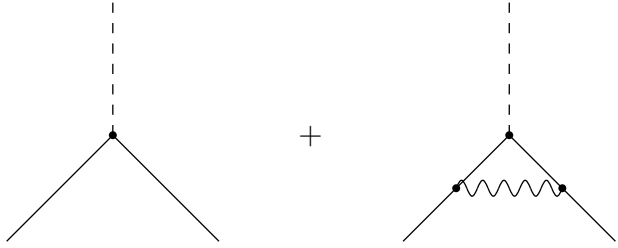}
\vskip -20 cm
\caption{Diagrams representing the starting guess for the vertex function in the iteration procedure described in the text.}
\label{fig:fig1}\end{center}
\end{figure}

To develop the PVC scheme within the present formalism a convenient starting ansatz for the vertex function is the one that includes explicitly the phonon degrees of freedom at the lowest level. This is depicted in Fig. \ref{fig:fig1} and its analytical expression reads
\beq
\begin{array}{ll}
\Gamma_{rh;ij}(t_4t_3;t_5) &\,=\, \delta_{ri}\delta_{hj}\delta(t_4 - t_5)\delta(t_3 - t_5) \\
 \ &\,  \\
&\,+\, \sum_{ls}\sum_{k_1k_2k_3k_4} <\,k_1\,\, r\, |\,\, f\,\, |\, k_2\,\, l\, >_A
\Pi_{k_2k_1;k_3k_4}(t_4,t_3)  <\,k_3\,\, s\, |\,\, f\,\, |\, k_4\,\, h\, >_A G_{li}(t_4,t_5) 
G_{js}(t_5,t_3) \\ 
\end{array}\label{eq:gamap}\eeq
To see the connection of this ansatz with the corresponding one for the self-energy, we substitute Eq. (\ref{eq:gamap}) for $\Gamma$ in Eq. (\ref{eq:Mpri}) to get the expression for $\tilde{M}$, according to Eq. (\ref{eq:Mtil}). Let us first consider the terms with at most one phonon
\beq
\begin{array}{ll}
&\tilde{M}_{ph}(t_1,t_3) \,=\, \frac{i}{2} \sum_{nr } <\, r\,\, p\, |\,\, f\,\, |\, h\,\, n\, >_A
G_{nr}(t_1,t_1^+) \delta(t_1 - t_3) \,+  \\
\  &\, \\
 &\ \,+\,  \frac{1}{2} \sum_{nr}\sum_{k_1k_2k_3k_4}\Big(\Big[ 
  \sum_{ij} i <\, i\,\, p\, |\,\, f\,\, |\, j\,\, n\, >_A
 \sum_{ls}
 <\, k_1\,\, r\, |\,\, f\,\, |\, k_2\,\, l\, >_A \Pi_{k_2k_1;k_3k_4}(\overline{t_4},t_3)\times \\
\  &\,  \\  
 &\, \ \ \ \ \ \ \ \ \ \ \ \ \ \ \ \ \ \times <\, k_3\,\, s\, |\,\, f\,\, |\, k_4\,\, h\, >_A G_{li}(\overline{t_4},t_1)G_{nr}(t_1,\overline{t_4})  \Big]  G_{js}(t_1,t_3) \,+ \\
\  &\,  \\
  &\ \,+\,  <\, k_1\,\, p\, |\,\, f\,\, |\, k_2\,\, n\, >_A 
\Pi_{k_2k_1k_3k_4}(t_1,t_3) <\, k_3\,\, r\, |\,\, f\,\, |\, k_4\,\, h\, >_A G_{nr}(t_1,t_3) 
\Big)  \,+ \\
\  &\, \\
&\ \,+\, \frac{1}{2} \sum_{nr} <\, n\,\, p\, |\,\, f\,\, |\, h\,\, r\, >_A 
< \psi_n^\dag(t_1) \psi_r(t_1) > \delta(t_1 - t_3)  \\     
\end{array}
\label{eq:Mapp}\eeq
As already anticipated, the first and fourth terms on the right hand side cancel out, and only the two terms with one phonon appears in the expression.   
In the first of these two terms, inside the square brackets a factor appears that can be schematically indicated as of the type $ G G  f \Pi$. It is then possible to use the general equation (\ref{eq:pigam}) for $\Pi$, with $\Gamma$ given by the first term only (delta function), which corresponds to the RPA approximation. Schematically
the relation $G G f \Pi \,=\, -\Pi + G G$ holds true, and explicitly
\beq
\begin{array}{ll}
\ &i \sum_{k_1k_2rl} G_{nr}(t_1,\overline{t_4}) G_{li}(\overline{t_4},t_1) 
<\, k_1\, r\, |\,\, f\,\, |\, k_2\,\, l\, >_A \Pi_{k_2k_1;k_3k_4}(\overline{t_4},t_3)
 \,=\, \\
\ &\, \\
\ &\ \,=\, -\Pi_{in;k_3k_4}(t_1,t_3) \,+\, G_{nk_3}(t_1,t_3) G_{k_4i}(t_3,t_1) \\
\end{array}    
\label{eq:ggp}\eeq
\par
\noindent
Once this expression is inserted in Eq. (\ref{eq:Mapp}), the first term on the right hand side of Eq. (\ref{eq:ggp}) sums up with an identical term appearing in that expression, and finally for $\tilde{M}$
one gets
\beq
\begin{array}{ll}
\tilde{M}_{ph}(t_1,t_3) &\,=\, \sum_{nrk_1k_2k_3k_4} <\, k_1\,\, p\, |\,\, f\,\, |\, k_2\,\, n\, >_A 
  \Pi_{k_2k_1k_3k_4}(t_1,t_3) <\, k_3\,\, r\, |\,\, f\,\, |\, k_4\,\, h\, >_A G_{nr}(t_1,t_3) \\
\ &\, \\
 &\,-\, \frac{1}{2} \sum_{nsijk_3k_4} <\, i\,\, p\, |\,\, f\,\, |\, n\,\, j\, >_A
 G_{nk_3}(t_1,t_3) G_{k_4i}(t_3,t_1) <\, k_3\,\, s\, |\,\, f\,\, |\, k_4\,\, h\, >_A
 G_{js}(t_1,t_3)   
\end{array}\label{eq:Mappf}\eeq      
\par
\begin{figure}[b]
%
\begin{center}
\includegraphics[bb= 240 0 400 790,angle=0,scale=1.]{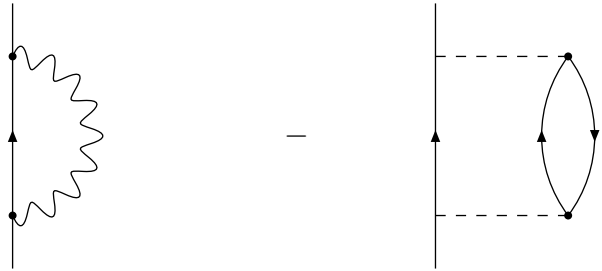}
\vskip -15 cm
\caption{First order diagrams for the self-energy beyond HF, consistently with the ansatz for $\Gamma$ in Fig. 1. Notice the subtraction of the ''bubble diagram" : its symmetry factor and other details are discussed in the text. }
\label{fig:fig2}\end{center}
\end{figure}
\begin{figure}[b]
\vskip 8 cm
\begin{center}
\includegraphics[bb= 240 0 400 790,angle=0,scale=1.]{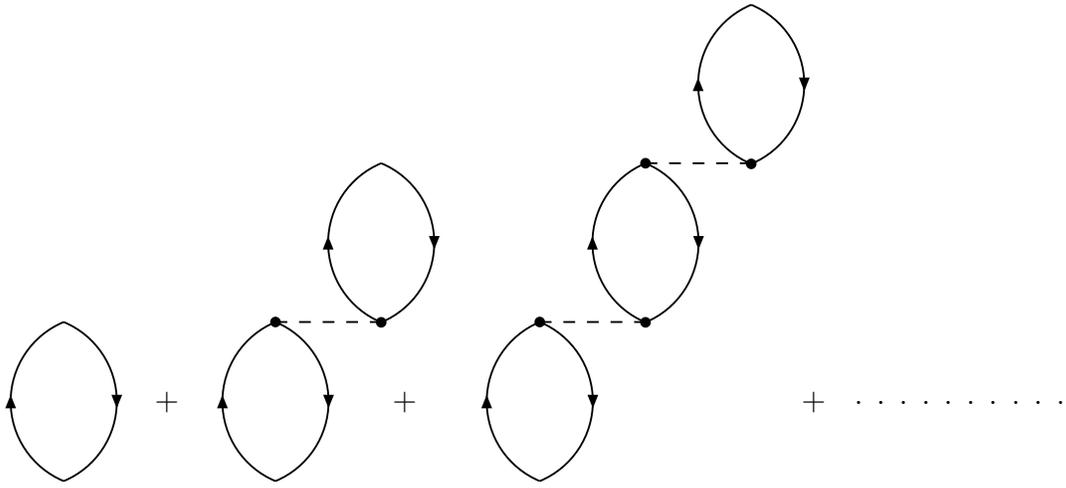}
\vskip -22 cm
\caption{The diagram series for the RPA approximation. }
\label{fig:fig3}\end{center}
\end{figure}

The corresponding diagrams are depicted in Fig. \ref{fig:fig2}. The reason why it is necessary to subtract  the one-bubble diagram from the one-phonon one has been discussed in previous works, see ref. \cite{Vin} and the review paper \cite{CM}. If one considers that the phonon in the RPA approximation is the sum of many-bubbles diagrams, see Fig. \ref{fig:fig3}, one notices that in the one-phonon diagram for the self-energy $\tilde{M}$, as reported in Fig. \ref{fig:fig2}, the first (one bubble) diagram contains two equivalent lines, and therefore the symmetry factor $1/2$ is necessary. This is not the case for all the diagrams with more than one bubble, where no equivalent lines appear, and therefore the symmetry factor 1 has to be used. The subtraction of the one bubble diagram to the one-phonon diagram produces then the correct symmetry factor for the first term in the RPA series.\par  
The explicit expression of the first term in Eq. (\ref{eq:Mappf}), which includes 
the particle-vibration coupling, can be obtained in general by Fourier transforming to the energy representation and expanding both the single particle Green's function and the phonon propagator in their spectral representation. Assuming for simplicity that the Green's function is diagonal,
one gets
\begin{equation}
\begin{array}{ll}
M_{q}(\omega) &\,=\, \sum_{n \{k\}} <\, k_1\,\, q\, |\,\, f\,\, |\, k_2\,\, n\, >_A
 \Big[\,\, \sum_{\lambda\alpha} 
 \frac{<\,\Psi_0 | \rho_{k_1 k_2}^\dag | \lambda\,>\,<\, \lambda | \rho_{k_3 k_4} | \Psi_0\,> 
 |<\,\alpha | \psi_n^\dag | \Psi_0\,>|^2}{\omega - \omega_\lambda - \omega_\alpha + i\eta} \\
 \ \,& \\
 \  &\,\,\,\,+\, \sum_{\lambda\beta} 
 \frac{<\,\Psi_0 | \rho_{k_3 k_4} | \lambda\,>\,<\, \lambda | \rho_{k_1 k_2}^\dag | \Psi_0\,> 
 |<\,\beta | \psi_n | \Psi_0\,>|^2}{\omega + \omega_\lambda + \omega_\beta - i\eta}\,\,\,\, \Big] 
 <\, k_3\,\, n\, |\,\, f\,\, |\, k_4\,\, q\, >_A  \\ 
\end{array}
\label{eq:spectr}\end{equation}
\noindent where $\lambda$ labels the excited vibrational states with energy $\omega_\lambda$ of the nucleus with atomic number A, and $\alpha, \beta$ the states of the $A+1, A-1$ systems with energy $\omega_\alpha, \omega_\beta$,
respectively. The states $\alpha, \beta$ coincide with the HF particle and hole states, and the expression of Eq. (\ref{eq:spectr})
can be written
\begin{equation}
M_q(\omega) \,=\, \sum_{p \lambda} \frac{| F_{qp}^\lambda |^2}{\omega - \omega_\lambda - \omega_p + i\eta} \,\,+\,\,
 \sum_{h \lambda} \frac{| F_{qh}^\lambda |^2}{\omega + \omega_\lambda + \omega_h - i\eta}   
\label{eq:F}\end{equation}
\noindent where $p$ and $h$ label particle and hole states, respectively, and
\begin{equation}
F_{qp}^\lambda \,=\, \sum_{k_1 k_2} \, <\,\Psi_0\, | \rho_{k_1 k_2}^\dag |\, \lambda\, >  
\,<\, k_1\,\, q\, |\,\, f\,\, |\, k_2\,\, p\, >_A  
\label{eq:Famp}\end{equation}
\noindent and analogously for $F_{hq}$. 
\par
Since the excited states $\lambda$ are described in the RPA scheme (the starting approximation for our iteration procedure), one recovers the standard expression of the diagram, see ref. \cite{Colo2010},
because in this case the matrix elements $ <\,\Psi_0\, | \rho_{k_1 k_2}^\dag |\, \lambda\, > $
can be identified with the amplitudes $X^\lambda_{ph}$ and $Y^\lambda_{ph}$ \cite{RS}. 
In ref. \cite{retPeter} it was shown that taking for $\omega$ the unperturbed value $\omega_q$ 
in the expression for $M_q(\omega)$ can be a good approximation.
\par
\begin{figure}[b]
\vskip 8 cm
\begin{center}
\includegraphics[bb= 280 0 440 790,angle=0,scale=1.]{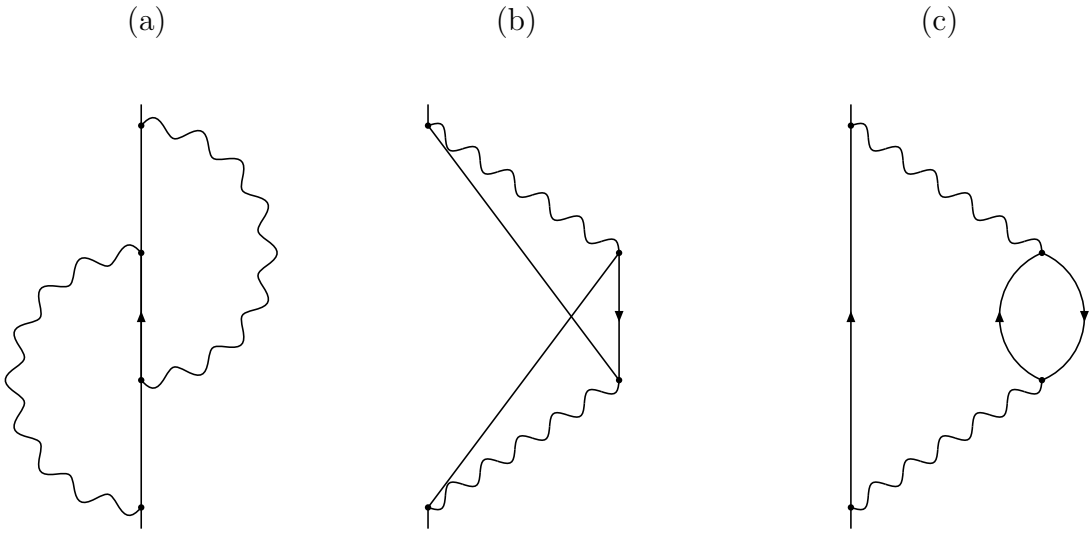}
\vskip -21 cm
\caption{(a) (b) are two-phonon diagrams arising in the self-energy from the ansatz for $\Gamma$ in Fig. 1. For details and meaning of diagrams (b) and (c), see the text.}
\label{fig:fig4}\end{center}
\end{figure}

    
As already noticed, two identical terms appear in the expression (\ref{eq:Mapp}), each one with a factor $1/2$, that sum up to the final result (\ref{eq:Mappf}). 
 Besides these two terms, one obtains also two-phonon terms for the self-energy. Their graphical representations are reported in Fig. \ref{fig:fig4}. Other time-ordering are possible, but we limit the discussion to these two cases. The origin and meaning of the diagram (b) can be understood looking at the inner structure of the phonon. In fact, if one considers the RPA series that defines the phonon, see Fig. \ref{fig:fig3}, inserted in the 
diagram of Fig. \ref{fig:fig2}, each bubble contains a particle (hole) line that can have the same quantum number as the particle (hole) coupled to the phonon, as illustrated in Fig. \ref{fig:fig4}c, where a bubble of the phonon is put in evidence.
To keep the Pauli principle correctly one has to consider also the diagram obtained by exchanging these two lines, which is indeed diagram \ref{fig:fig4}b.  This diagram is therefore demanded by Pauli principle. 
All the above discussion shows how the expansion is able to take into account Pauli principle and to include the correct symmetry factor, which is not trivial once the phonon degrees of freedom are explicitly introduced at microscopic level. 
\par       
 We notice that the choice of Eq. (\ref{eq:gamap}) for the $\Gamma$ function modifies the integral equation for the propagator $\Pi$. It is useful to introduce the polarization propagator $P$, which is defined in the present formalism as
\beq
 P_{mk;ij}(t,t_3) \,=\, \sum_{pq} G_{mp}(t,\overline{t_1}) G_{qk}(\overline{t_2},t)\Gamma_{pq ; ij}
(\overline{t_1},\overline{t_2} ; t_3)
\eeq
\noindent
It appears as a kernel of the integral Eq. (\ref{eq:pigam}) for the phonon propagator $\Pi$, which in fact can be written
\beq
\Pi_{mk , rs}(t,t') =\,\, \sum_{ij} P_{mk ; ij}
(t,\overline{t_3})
\big[\, \delta_{ir}\delta_{js}\delta(\overline{t_3} - t') - i \sum_{tu}  <\, i\,\, t\, |\,\, f\,\, |\, j\,\,u\,>_A \Pi_{ut , rs}(\overline{t_3},t')\, \big]  
\label{eq:pipol}\eeq
\par
\begin{figure}[b]
\vskip 8 cm
\begin{center}
\includegraphics[bb= 240 0 400 790,angle=0,scale=1.]{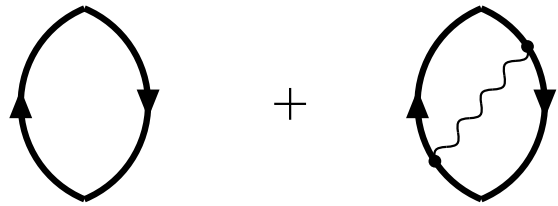}
\vskip -23 cm
\caption{Polarization diagrams as obtained at the second iteration step. The thick lines indicate that the Green' s function are considered consistently beyond the HF approximation. }
\label{fig:fig5}\end{center}
\end{figure}

\noindent  
The polarization propagator corresponding to this choice for the $\Gamma$ function is 
depicted in Fig. \ref{fig:fig5}. The two diagrams (a) and (b) introduce an implementation of the RPA approximation. Notice that the single particle Green' s function $G$ has to be calculated with the self-energy just discussed. This is indicated by the thick lines that represent the propagator $G$.
In the sequel the self-consistency of the Green' s function will be assumed in any case. Diagram (b) includes the coupling of one particle-one hole excitations to two particle-two holes configurations, described as one phonon plus a one particle-one hole excitations. In the present scheme the two correlation effects, on the self-energy and on the phonon propagator, have to be taken together. This prescription appears as a consequence of the conserving structure of the basic equations, as discussed in Section 2. These effects have been considered at perturbative level  
in the study of the Giant Resonances widths in reference \cite{RMP}, and, using Skyrme forces, in refs. \cite{Gian1,Gian2,Gian3}. 
A similar scheme has been introduced in ref. \cite{Edu,RiLi}, and applied within the
{\it Time-Blocking Approximation} \cite{TBA}.\par 
It has to be noticed that the structure of the diagrams of Fig. \ref{fig:fig5} includes the coupling of particle-hole states to two particle-two hole states. As such it is physically equivalent to the second RPA \cite{DaP,Gamba} for the phonon excitations. In the present scheme the equation for the phonon can be written in the RPA form, but with the addition of an energy dependent effective force \cite{DaP}, that comes from the virtual excitation of two particle-two hole states. A detailed comparison of this formulation of the second RPA with others \cite{DaP,Gamba} is outside the scope of the present work. We only notice that the single particle self-energy modifies also the density matrix and consequently, as already stressed, the static single particle potential through the one-body interaction $U$ of Eq. (\ref{eq:Upot}). 
It has also to be noticed that these modifications on the phonon can be seen as the introduction of anharmonicity in the vibrational modes due to their couplings. 
\par
\begin{figure}[b]
\vskip 8 cm
\begin{center}
\includegraphics[bb= 240 0 400 790,angle=0,scale=1.]{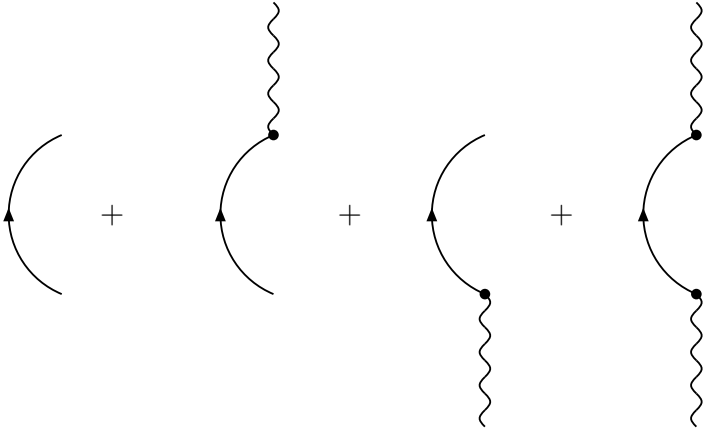}
\vskip -20 cm
\caption{Pictorial description of the effect of the functional derivative on the phonon propagator. The first picture on the left corresponds to the functional derivative of $\Pi^0$, which appears at the left hand side of Eq. (\ref{eq:DPG}). See the text for further explanation.}
\label{fig:fig6}\end{center}
\end{figure}

      
 A second iteration can be performed, which will produce a set of other diagrams, 
  but we will not analyse in detail the additional terms generated in this way, since the applications will focus on the first order diagrams.
Here we consider a general problem that is encountered for the evaluation of the functional derivative of the self-energy. 
In general the derivative "cuts" a single particle line within a given diagram for $\tilde{M}$.
This line can appear explicitly in the diagram, but it can be also implicitly present in the phonon propagator. It is therefore essential to find a method to perform such a derivative "inside" the phonon. This can be done by displaying the expansion of the phonon propagator in terms of single-particle 
Green' s  functions and interaction matrix elements. Let us illustrate the procedure in the  case of the RPA approximation. In this case the phonon propagator is the sum of a series of diagrams with increasing number of "bubbles" , see Fig. \ref{fig:fig3}. The effect of the functional derivative can be figured out by realizing that the Green 's function to be "cut" can be present at the beginning of each diagram, at the end of it or in the middle. Once the cut has been performed, the remaining parts
of the diagram can be again summed up, and the phonon propagator is again constructed. If the derivative is in the middle of the diagram, on both sides of the cut the phonon can be reconstructed.
In the other cases this can be done only on one side. Furthermore for the first diagram with only one bubble no phonon can be reconstructed. The functional derivative of the phonon propagator can be then depicted graphically as in Fig. \ref{fig:fig6}. The corresponding analytical expression reads
\beq\begin{array}{ll}
 &i\frac{\delta}{\delta G_{rs}(t_6,t_7)} \left( \Pi_{hk;lm}(t_1,t_4) \,-\, \Pi^{(0)}_{hk;lm}(t_1,t_4) \right) \,=\, \ \ \ \ \ \ \ \ \ \ \ \ \ \ \ \ \ \ \ \ \ \ \ \ \ \ \ \ \ \ \ \ \ \ \ \ \ \ \ \ \ \  \\
\ &\,   \\ 
\ &\,=\, \sum_{pqtu} \frac{\delta \Pi^{(0)}_{hk;pq}(t_1,\overline{t_2})}{\delta G_{rs}(t_6,t_7)}
 <\, p\,\, t\, |\,\, f\,\, |\, q\,\, u\, >_A  \Pi_{ut;lm}(\overline{t_2},t_4) \,+\,   \\
\ &\,   \\
\ &\,+\, \sum_{pqtu} \Pi_{hk:pq}(t_1,\overline{t_2}) <\, p\,\, t\, |\,\, f\,\, |\, q\,\, u\, >_A
\frac{\delta \Pi^{(0)}_{ut;lm}(\overline{t_2},t_4)}{\delta G_{rs}(t_6,t_7)} \,+\,  \\
\ &\, \\
\ &\,+\, \sum_{ijab} \Pi_{jk;ab}(t_1,\overline{t_3}) <\, a\,\, i\, |\,\, f\,\, |\, b\,\, j\, >_A                                                   
\frac{\delta \Pi^{(0)}_{ji;pq}(\overline{t_3},\overline{t_2})}{\delta G_{rs}(t_6,t_7)}                    <\, p\,\, t\, |\,\, f\,\, |\, q\,\, u\, >_A \Pi_{ut;lm}(\overline{t_2},t_4)  \\
\end{array}\label{eq:DPG}\eeq
\noindent
where $\Pi^{(0)}$ is the free polarization propagator, and its functional derivative corresponds to the first diagram of Fig. \ref{fig:fig6}. This result is in line with the analogous
expression derived in reference \cite{Baym}.
The functional derivative of the phonon propagator is essential if the approach has to be kept at a fully microscopic level. 
\par
\section{Density dependent forces \label{sec:ddep}}
The diagrammatic expansion introduced in the previous sections implements the formalism of the Nuclear Field Theory
\cite{Mott,PFBPR,NPA260} to the case of a realistic effective two-body forces on a microscopic basis. Further developments are necessary if modern effective forces or functionals are considered.    
 In general the effective force depends on the
density or the density matrix and the formalism must be generalized to this case. In minimizing the EDF one has to consider the functional variation of the force, and the mean field (Hartree-Fock) potential $\cal V$ does not involve only the effective force but also its functional derivative
\beq
\begin{array}{ll} 
%
 < k\,\, |\,\, {\cal V}\,\, |\,\, k' > &\,=\, \big( \frac{\delta V}{ \delta \rho_{kk'}}\big)_0 \\
 \  &\,        \\
 \  &\,=\, \frac{1}{2} \sum_{lj} \big[ f_{kl ; k'j}^A \rho^0_{lj}  \,+\, \rho^0_{lj} f_{lk ; jk'}^A \,+\,
 \sum_{ip} \rho^0_{ip}\big( \frac{\delta f_{il ;pj}^A }{ \delta \rho_{kk'} }\big) \rho^0_{lj} \big] \\
\end{array}\label{eq:VHF}\eeq
\noindent
 Solving self-consistently the resulting Hartree-Fock equations with this effective single particle potential is equivalent to minimize the full functional  of Eq. (\ref{eq:fun}). Notice that the first two terms in square brackets are equal.  
\par 
The dynamics of the density fluctuations is determined by the energy surface around the minimum.
If the fluctuations are not too large, one can expand the functional to second order in the fluctuations
$ \delta \rho_{ij} $ 
%
\beq
E \,=\, E_0 \,+\, \Big( \frac{\delta V}{ \delta \rho_{ij}}\Big)_0 \delta \rho_{ij} \,+\,
\frac{1}{2} \Big( \frac{\delta^2 V }{ \delta \rho_{st}\delta \rho_{uv} }\Big)_0 \delta \rho_{st} \delta \rho_{uv}
\eeq 
Also in calculating the second derivative  one has to take into account the density matrix dependence of the effective force
\beq
\begin{array}{ll}
%
 < \,u\,\, s\, |\,\, \kappa\,\, | \,v\,\, t\, >_A &\,=\,
 \big( \frac{\delta^2 V }{ \delta \rho_{st}\delta \rho_{uv} }\big)_0 \\
 \  &\,        \\
 \  &\,=\, \frac{1}{2} \big\{\ f_{us ; vt}^A  \,+\, f_{su ; tv}^A \,+\,
  \big[  \sum_{lj}\big(\ \rho^0_{lj} \big( \frac{\delta f_{lu ; jv}^A }{ \delta \rho_{st} } \big)\,+\,
 \big( \frac{\delta f_{sl ; tj}^A }{ \delta \rho_{uv} }
 \big) \rho^0_{lj}\ \big)  \,+\,
 \sum_{iklj} \rho^0_{ik} \big( \frac{\delta^2 f_{il ; kj}^A}{ \delta \rho_{st}\delta \rho_{uv}
  }\big) \rho^0_{lj}\ \big]\big\} \\
\end{array}\label{eq:vff}\eeq
\noindent 
The quantity $ \kappa $ can be interpreted as the interaction between two density fluctuations. 
This is a well known result, usually considered in connection with the standard RPA approximation for the vibrational states  \cite{RS}, and we report it here for later discussion. As such Eq. (\ref{eq:vff})
gives the matrix elements of the basic interaction between two elementary particle-hole excitations.
If the effective force is density independent, only the first two terms in the curly brackets are present and the matrix element, as expected, is correctly anti-symmetrized and it coincides with the previous two-body matrix element in the case of density independent forces (notice that the first two terms are actually equal). The density dependence introduces the additional "rearrangements" terms involving functional derivatives. These terms are not automatically anti-symmetrized, despite we started with the anti-symmetrized matrix elements of
 Eq. (\ref{eq:antif}). This is due to the already mentioned lack of anti-symmetry with respect to orbitals that appear in the density dependence of the effective force. In principle one should then anti-symmetrize
these matrix elements. This is a general problem for the density dependent effective forces. 
Relying on the first line of Eqs. (\ref{eq:VHF}) and (\ref{eq:vff}), we shall
develop the formalism for a generic energy density
functional. In that case, we are not supposed to trace back
any direct or exchange term of any starting interaction $V$.
Of course, if such interaction exists like in the case of
the Gogny force and most of the Skyrme sets, the equations
on the second line of Eqs. (\ref{eq:VHF}) or (\ref{eq:vff}) can be of some help.
The identification of the residual interaction with the second functional derivative 
can be further clarified if the density dependence of the functional is considered explicitly. The single particle density $ \rho (\mathbf{r}) $ can be written in terms of the density matrix (neglecting spin-isospin labels)
\beq
    \rho (\mathbf{r}) \,=\, <\, \psi(\rv)^\dag\, \psi(\rv)\, > \,=\, \sum_{ij} <\, \psi_i^\dag\, \psi_j\, >\, \phi_i(\rv)^* \phi_j(\rv)  \,=\, \sum_{ij} \rho_{ij}\, \phi_i(\rv)^* \phi_j(\rv) 
\label{eq:denm}
\eeq
\noindent
\noindent For a fixed single particle basis $ \{ \phi \} $, the variation of the density matrix implies a definite variation of the density. It follows that the functional derivative of $ V $ with respect to the density matrix can be 
written
\beq
\frac{\delta V}{\delta \rho_{ij}} \,=\, \int d^3r \frac{\delta V}{\delta \rho(\rv)}\, \frac{\partial \rho(\rv)}{\partial \rho_{ij}} \,=\, \int d^3r \frac{\delta V}{\delta \rho(\rv)}\, \phi_i(\rv)^*\, \phi_j(\rv) \,=\, <\, i\, | \frac{\delta V}{\delta \rho(\rv)} |\, j\, > \,\equiv\, < i\,\, |\,\, {\cal V}\,\, |\,\, j >  
\label{eq:derV}
\eeq
\noindent Similarly one gets
\beq
\frac{\delta^2 V}{\delta \rho_{ij} \delta \rho_{lm}} \,=\, \int d^3r\int d^3r' \frac{\delta^2 V}{\delta \rho(\rv) \delta \rho(\rv')} \phi_i(\rv)^*\, \phi_j(\rv)\, \phi_l(\rv')^*\, \phi_m(\rv') \,\equiv\, < \,i\,\, l\, |\,\, \kappa\,\, | \,j\,\, m\, >  
\label{eq:derV2}
\eeq
\par\noindent
This is the general form of the effective density-density interaction associated with a given EDF if one can stick to the harmonic approximation, i.e., 
stop the expansion of the potential with respect to the density at second order. 
This can be justified in specific cases by looking at the transition density associated with the vibrational states, which can be small with respect to the corresponding unperturbed density, see e.g. figure 1 of ref. \cite{Iku}.  
\par
The harmonic approximation, i.e., the truncation of the expansion at the second order,  may be inappropriate
in some cases; however, one could in the same way as
in Eqs. (\ref{eq:derV},\ref{eq:derV2}) identify the matrix elements associated
with third, fourth, etc. order variations of the potential.
\par
At this point the development of the formalism can follow exactly the same 
lines as in the case of a density independent force.  
We adopt the same interaction that describes the density-density coupling also for the particle-vibration
coupling. This goes along the line of Ref. \cite{Waro}, 
where it is shown that the matrix elements for the 3 particle-1 hole 
coupling processes have indeed the form of Eq. (\ref{eq:derV2})
(in all cases, except for the quantum numbers $J^\pi \,=\, 0^+$ of the excited particle-hole).
The problem of double counting, discussed previously, arises 
also in this case, and again a refitting of the force parameters 
should be envisaged.  

However, an additional caveat has to be considered.
Since the equations of motion are derived using the form of Eq. (\ref{eq:vff}) for the matrix elements of the two-body interaction,   
the processes that determine the single particle dynamics 
and the nuclear vibrations can include particle-particle scattering, 
for which the density-density interaction matrix elements of 
Eq. (\ref{eq:vff}) does not look appropriate. In these 
processes, in fact, the particle-particle or hole-hole interaction 
does not involve density fluctuations \cite{Waro}. The hamiltonian of 
Eq. (\ref{eq:ham}) necessarily produces these elementary processes. 
This feature is intrinsic to the hamiltonian formalism, where 
the two-body matrix elements are fixed from the start and they 
cannot depend on the processes where they are involved. 
\par
\begin{figure}[b]
%
\begin{center}
\includegraphics[bb= 240 0 400 790,angle=0,scale=1.]{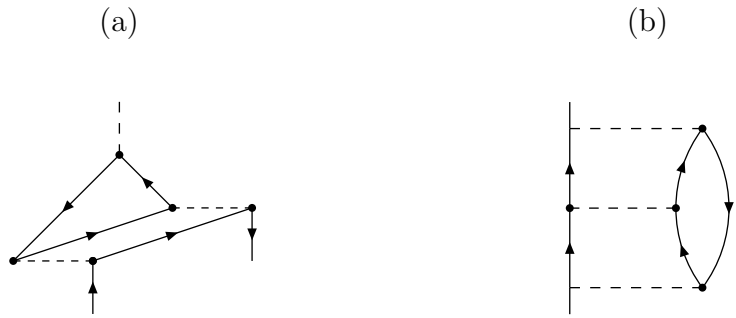}
\vskip -20 cm
\caption{(a) Diagrams for the vertex function generated by the functional derivative of the second diagram of Fig. (\ref{fig:fig2}).(b)The corresponding self-energy diagram, where the interaction in the middle corresponds to a particle-particle scattering process, see the text. }
\label{fig:fig7}\end{center}
\end{figure}

  
To see how particle-particle scattering arises in the present formalism, 
let us consider the functional derivative of the second diagram 
(the ''bubble'') of Fig. \ref{fig:fig2}, which is performed at the second iteration of the expansion procedure. If the functional derivative acts on the hole line of the diagram, the vertex function develops the diagram (a) of Fig. (\ref{fig:fig7}), which in turn generates the diagram (b) for the self-energy. The interaction at the center of this diagram is indeed
a particle-particle scattering, which is expected to be described by a matrix element different from (\ref{eq:derV2}).  
\par
The particle-vibration model  assumes that this processes have negligible effects, since the particle-vibration coupling and the particle-hole matrix elements are considered the dominant ones. In this case no conceptual difficulty can arise. Along the same talking, the approximations obtained in the expansion are conserving to the extent that these processes can be neglected.\par 
In any case we will restrict the applications to the lowest order approximation, where this problem is absent, and we will leave a throughout analysis of this point to future works. 
\vskip 0.4 cm
\section{Application to $^{40}Ca$. \label{sec:app}}

The lowest order dynamical contribution to the single particle self-energy is given by the two diagrams of Fig. \ref{fig:fig2}. 
Notice that they have to be subtracted. As discussed in the derivation of this result in Sect. III A, the subtraction of 
the bubble diagram is needed once the phonon degrees of freedom is introduced in order to fix the correct symmetry factor. 
We will study the relevance of
this additional diagram in the specific case of the nucleus $^{40}$Ca.
 It should be noted that this additional diagram had been considered
in the past, but not in any of the recent self-consistent PVC calculations.
Our study, therefore, is the first of such kind with the consistent
use of the whole Skyrme central force.
\par 
Let us first consider the case of the density independent Skyrme force SV \cite{SV}. In Table I we report 
the single particle energy levels around the Fermi energy. 
The Hartree-Fock (HF) energies are displayed, together with the 
phonon diagram contribution $E_{ph}$ and the bubble diagram contribution 
$E_{bub}$. The total correction $E_{cor} \equiv E_{ph} - E_{bub}$ is 
added to the 
HF energy to obtain the ``dressed'' single particle energies $E$. 
The calculations have been performed by simply setting an upper cut-off 
of the single-particle energies at 60 MeV: the RPA calculations and PVC
calculations are done within this model space and no additional selection
has been made, so that in particular all RPA phonons with multipolarity 
0$^+$, 1$^-$, 2$^+$, 3$^-$ and 4$^+$ are retained. 
In this way, the phonon contributions are corrected consistently by 
the corresponding bubble diagrams. 
The self-energy corrections due to the PVC coupling turn out to be all negative, both for particle and hole states. The reason of this particular feature is discussed in ref. \cite{Colo2010}.
One can notice that the correction due to the bubble diagram is not negligible and in some cases it is substantial. 

\begin{table}
\caption{Single particle energy levels of $^{40}$Ca for the Skyrme force SV. The column $HF$ contains
the Hartree-Fock energies, $E_{ph}$ and $E_{bub}$  indicate the contribution of the one-phonon
diagram and the bubble diagram of Fig. \ref{fig:fig2}, respectively, $E_{cor}$ their difference.
The column labeled $E$ contains the corrected level energies and $E_{exp}$ the
experimental ones. Energies are in MeV. }
\begin{ruledtabular}
\begin{tabular}{ccccccc}

 $   $ & $HF$ & $E_{ph}$ & $E_{bub}$ &  $E_{cor}$  &  $E$  &  $E_{exp}$ \\
\hline
  1d5/2 & -27.67 & -0.76 &  0.53 & -1.28 & -28.95 & -22.39  \\
  2s1/2 & -19.34 & -4.25 & -0.42 & -3.83 & -23.17 & -18.19  \\
  1d3/2 & -18.95 & -1.58 &  0.32 & -1.90 & -20.85 & -15.64  \\
  1f7/2 &  -7.54 & -0.87 & -0.30 & -0.57 &  -8.11 &  -8.62  \\
  2p3/2 &  -2.02 & -2.28 & -0.75 & -1.53 &  -3.55 &  -6.76  \\
  2p1/2 &  -0.37 & -1.74 & -0.60 & -1.14 &  -1.51 &  -4.76  \\
  1f5/2 &   1.62 & -1.21 & -0.42 & -0.79 &   0.83 &  -3.38  \\
\end{tabular}
\end{ruledtabular}
\vskip -0.5 cm
\end{table}

\par 
To study the dependence of the results on the adopted effective force, we consider the more modern
Skyrme force Sly5 \cite{Sly5}. Since the force is density dependent, we follow the scheme described in Sec. IV.
The HF energies and the self-energy contributions are reported in Table II. 
Even in this case, one can notice that the bubble diagram is in general a substantial fraction 
of the phonon diagram contribution and its overall size is even larger than for the SV Skyrme force.
Therefore, we conclude that the inclusion of the bubble diagram certainly requires careful consideration 
in all future calculations performed in the PVC scheme.

We do not wish to discuss in detail the comparison with experimental data, let alone to claim 
that the inclusion of the PVC coupling is improving such comparison. We could note, on the one hand, that 
the r.m.s. devation with respect to experiment of the HF energies is $1.28$ MeV, and that this value increases 
to $2.40$ MeV with the inclusion of the one phonon diagram contribution $E_{ph}$ but it decreases to $0.78$ MeV 
with the addition of the bubble diagram contribution $E_{bub}$. On the other hand, one must be aware that 
the results obtained with a zero-range force are very sensitive to the choice of the model space and can
diverge if no cut-off is set. While we have started to tackle this problem \cite{Brenna,Rizzo}, the present
results are only meant to illustrate that the lowest-order approximation to PVC for single-particle states
must include the bubble diagram.

\par
\begin{table}
\caption{Single particle energy levels of $^{40}$Ca for the Skyrme force Sly5. Notation as in Table I. }
\begin{ruledtabular}
\begin{tabular}{ccccccc}

 $   $ & $HF$ & $E_{ph}$ & $E_{bub}$ &  $E_{cor}$  &  $E$  &  $E_{exp}$ \\
\hline
  1d5/2 & -22.09 & -1.30 & -0.88 & -0.42 & -22.51 & -22.39  \\
  2s1/2 & -17.27 & -3.54 & -2.29 & -1.25 & -18.52 & -18.19  \\
  1d3/2 & -15.19 & -1.65 & -0.98 & -0.67 & -15.86 & -15.64  \\
  1f7/2 &  -9.67 & -2.34 & -1.50 & -0.84 & -10.51 &  -8.62  \\
  2p3/2 &  -5.29 & -4.70 & -2.71 & -1.99 &  -7.28 &  -6.76  \\
  2p1/2 &  -3.10 & -4.64 & -2.72 & -1.92 &  -5.02 &  -4.76  \\
  1f5/2 &  -1.28 & -2.71 & -1.12 & -1.59 &  -2.87 &  -3.38  \\
\end{tabular}
\end{ruledtabular}
\vskip -0.2 cm
\end{table}

\section{Conclusions. \label{sec:conc}}
We have presented a many-body theory of nuclear structure within the 
particle-vibration coupling scheme, where the degrees of freedom are 
single particles and phonons, each one described by the corresponding 
propagator. The formalism is developed for a generic effective 
interaction or Energy Density Functional, which are supposed to 
be defined within a model space. Both density independent and density 
dependent effective interactions have been considered. The theory is 
devised to go beyond the mean field approximation to all orders. Since 
the effective interaction is expected to include part of the correlations, 
in principle a refitting of the effective force would be 
mandatory. We have not approached this problem, but instead we have 
focused on the correct treatment of the particle-vibration coupling 
when the phonon degrees of freedom are treated at a completely 
microscopic level.  
The method is based on the equations that couple the single particle 
self-energy, the vertex function and the phonon propagator, and 
are derived within the Hamiltonian formalism. The solution of the 
equations is by iteration. At each iteration one gets an approximation 
that is conserving \cite{BaymKad,Baym}, i.e. the vertex function 
conserves the scalar current and the total momentum. The terms 
so obtained by iteration can be represented by diagrams. In this way 
both static and dynamical
corrections beyond the mean field are systematically introduced. 
We must stress that the diagrammatic expansion takes into account 
systematically both the Pauli principle and the possible 
over-counting of contributions. 
 We have shown that with a simple one-phonon ansatz for the
vertex function one obtains in a natural way the lowest order
approximation for the self-energy in terms of
Fig. \ref{fig:fig2}. 
\par
The general expansion provides a formal extension of 
the Nuclear Field Theory \cite{Mott,NPA260,PFBPR} to realistic effective interactions   
and it is a non-perturbative solution of the many-body problem based on an effective interaction in a model space. Truncating the expansion at a given order provides a conserving approximation
within the particle-vibration coupling scheme. 

We have presented applications of the theory to the nucleus $^{40}$Ca. Both the density independent Skyrme 
force SV and the density dependent Sly5 were considered and compared. 
Already at this level the results indicate that it is essential to take into account correctly 
the microscopic structure of the phonon, which implies the introduction of the bubble diagram  of Fig. \ref{fig:fig2} to fix the correct symmetry factor for the underlying RPA diagrams that define the phonon propagator. The bubble diagram turns out to be a substantial correction to the one-phonon diagram, and demands a careful treatment in any application of the PVC coupling scheme.       

 In summary, we claim that even at the lowest order a sound PVC calculation should include
at least the two diagrams for the self-energy in Fig. 2. Although this result has been known
for some time, we are not aware of its implementation in recent 
self-consistent calculations 
of the self-energy with microscopic energy functionals (although such correction has been introduced
in the total energy calculations of Ref. \cite{Carlsson}). Our numerical
results indicate the importance of the bubble correction. More importantly, 
we pave the way to calculations that go beyond this lowest order approximation.

\appendix
\section{Derivation of the basic equations}
We give here some detail on the derivation of the set of the three equations that are the basis of the formalism
described in Section \ref{sec:form}. In general the equation of motion for $G$ can be written
\beq
\begin{array}{ll}
\frac{\partial }{ \partial t_1} G_{pq}(t_1,t_2) &\,=\, \delta(t_1 - t_2)\delta_{pq} + \sum_{p'}\epsilon_{pp'}
G_{p'q}(t_1,t_2) \\
 \ &\,  \\
&\,-\, i \frac{1}{2} \sum_{kmn} \big[ <\,k\,\, p\, |\,\, \kappa\,\, | \, m\,\, n\,>_A
<T\{\psi_k^\dag(t_1)\psi_m(t_1)\psi_n(t_1)\psi_q^\dag(t_2)\}>  \\
 \ &\,  \\
   &\,+\, <\,k\,\, p\, |\,\, \kappa\,\, | \, n\,\, m\, >_A  <\psi_k^\dag \psi_m>_0 G_{kq}(t_1,t_2)
   \big]
   + \sum_l \phi_{pl}(t_1) G_{lq}(t_1,t_2) \\
\end{array}\label{eq:mot}\eeq
\noindent where the label $A$ indicates anti-symmetrized matrix element. The two-body interaction term (second
line) is directly related to the functional derivative of the Green 's function with respect to the external
potential $\phi(t)$ \cite{BaymKad}
\beq
 -i <T\{\psi_k^\dag(t_1)\psi_m(t_1)\psi_n(t_1)\psi_q^\dag(t_2)\}> \,\,=\,\,
 i \frac{\delta G_{nq}(t_1,t_2) }{ \delta \phi_{km}(t_1)}
 \,+\, <\psi_k^\dag(t_1)\psi_m(t_1)> G_{nq} (t_1,t_2)
\label{eq:fder}\eeq
\noindent It has to be noticed that the mean-value at the right hand side is time dependent due to the external
interaction $\phi(t)$ and must be kept distinct from the static mean-values that appear in the expansion of the
normal product of Eq.(\ref{eq:np}). The former is just the time dependent one-body density matrix of the system.
Let us introduce the inverse Green 's function $G^{-1}$
\beq
 \sum_r G^{-1}_{pr}(t_1,\overline{t_2}) G_{rq}(\overline{t_2},t_1') \,=\, \sum_r
 G_{pr}(t_1,\overline{t_2})G^{-1}_{rq}(\overline{t_2},t_1') \,=\, \delta_{pq} \delta(t_1 \,-\, t_1')
\label{eq:Ginv}\eeq
\noindent where a line over a time variable indicates integration over that variable following the notation
already introduced in the text. From Eq. (\ref{eq:Ginv}) one gets
\beq
\frac{\delta G_{nq}(t_1,t_2) }{ \delta \phi_{km}(t_1)} \,=\, - \sum_{rs} G_{nr}(t_1,\overline{t_3}) \frac{\delta
G^{-1}_{rs}(\overline{t_3},\overline{t_4}) }{ \delta \phi_{km}(t_1)} G_{sq}(\overline{t_4},t_2)
\label{eq:fdinv}\eeq
\noindent If we insert this expression in Eqs. (\ref{eq:fder},\ref{eq:mot}), the equation of motion (\ref{eq:mot})
can be written according to Eqs. (\ref{eq:Dyson}-\ref{eq:Mtil}), where
\beq
 M_{ph}'(t_1,t_4) \,=\, - i \frac{1}{2}\sum_{kmnr} <\, k\,\, p\, |\,\, \kappa\,\, |\, m\,\, n\,> G_{nr}(t_1,\overline{t_3})
 \frac{ \delta G^{-1}_{rh}(\overline{t_3},t_4) }{ \delta \phi_{km}(t_1) }
\label{eq:self1}\eeq
By comparing Eqs. (\ref{eq:mot}) and (\ref{eq:fdinv}) one gets the expression of the inverse Green 's function
$G^{-1}$ in terms of the self-energy
\beq
\begin{array}{ll}
G^{-1}_{pq} \, =&\, i \frac{\partial }{ \partial t_1} \delta(t_1 \,-\, t_2) \, -\, \epsilon_{pq} \delta(t_1 \,-\, t_2) \\
\ &\,  \\
 \,&-\, \big[ M_{pq}'(t_1,t_2) \, +\, \frac{1}{2}\sum_{km} <\, k\,\, p\, |\,\, \kappa\,\, |\, q\,\, m\,>_A
 < \psi_k^\dag(t_1) \psi_m(t_1) > \delta (t_1 \,-\, t_2) \big]  \\
 \ &\, \\
 \,&-\, U_{pq}(t_1) \delta(t_1 \,-\, t_2)  \\
\end{array}\label{eq:Gm1}\eeq
\noindent where the potential $U$ is given by Eq. (\ref{eq:Upot}) and the quantity in square parenthesis is just
$\tilde{M}$ of Eq. (\ref{eq:Mtil}).

\par
The phonon degree of freedom can be introduced by noticing that the functional derivative of the potential $U$ is
related to the density-density response function, that can be identified with the phonon propagator. In fact the
functional derivative of $U$ can be computed according to the general method as in Eq. (\ref{eq:fder})
\beq
\frac{\delta U_{ph}(t) }{ \delta \phi_{rs}(t')} \, =\, \delta_{pr}\delta_{hs}\delta(t \,-\, t')
\, -\, i \sum_{km} <\, k\,\, p\, |\,\, \kappa\,\, |\, m\,\, h\,>_A \Pi_{sr ; km}(t,t')
\label{eq:phon}\eeq
\noindent where $\Pi$ is defined according to Eqs. (\ref{eq:pol}). In fact, according to the general result of Eq.
(\ref{eq:fder}), one has
\beq\begin{array}{ll}
 \frac{\delta <\psi_q^\dag(t_2)\psi_n(t_2)> }{ \delta \phi_{km}(t_1)} &\,\,=\,\,
i <T\{\psi_k^\dag(t_1)\psi_m(t_1)\psi_q^\dag(t_2)\psi_n(t_2)\}>
 \,-\, i <\psi_k^\dag(t_1)\psi_m(t_1)> <\psi_q^\dag(t_2)\psi_n(t_2)> \\
\, &\,  \\
 \ &\,\,\equiv\,\, -i \Pi_{mk;qn}(t_1,t_2)
\end{array}\label{eq:Uder}\eeq
%
%
%
%
\noindent  The particle-vibration coupling can be introduced using the chain property of the functional
derivative, since the potential $U$ is a functional of the external perturbation $\phi$. Indeed the $M'$
self-energy can be expressed in terms of the phonon degree of freedom
\beq\begin{array}{ll}
M_{ph}'(t_1,t_4) &\,=\, -i \frac{1}{2} \sum_{kmnrij} <\, k\,\, p\, |\,\, \kappa\,\, |\, m\,\, n\,>_A G_{nr}(t_1,\overline{t_4})
\big( \frac{\delta G_{rh}^{-l}(\overline{t_3},\overline{t_4}) }{ \delta U_{ij}(\overline{t_5})}\big) \frac{\delta
U_{ij}(\overline{t_5})}{ \delta
\phi_{km} (t_1)}  \\
\ &\,  \\
 &\,=\, \frac{1}{2} i \sum_{nrij} W_{pn ; ij}(t_1,\overline{t_5}) G_{nr}(t_1,\overline{t_3}) \Gamma_{rh ; ij}(\overline{t_3},
 t_4 ; \overline{t_5}) \\
\end{array}\label{eq:selfchi}\eeq
\noindent where
\beq\begin{array}{ll}
W_{pn ; ij}(t_1,t_5) &\,=\, \sum_{km} <\, k\,\, p\, |\,\, \kappa\,\, |\, m\,\, n\,> \big(\frac{\delta
U_{ij}(t_5)}{ \delta \phi_{km} (t_1)}\big)  \\
%
 \ &\, \\
\Gamma_{rh ; ij}(t_3,t_4 ; t_5) &\, =\, - \big(\frac{\delta G_{rh}^{-l}(t_3,t_4) }{ \delta U_{ij}(t_5)}\big)
\end{array}\label{eq:wgam}\eeq
\noindent Together with Eq. (\ref{eq:phon}), this gives Eqs. (\ref{eq:Mpri},\ref{eq:W}). The equations
(\ref{eq:pigam}) for $\Pi$ and $\Gamma$ can be derived using again the chain property of the functional
derivative. For the phonon propagator one gets
\beq\begin{array}{ll}
\Pi_{mk;qn}(t_1,t_2) &\,\, =\,\, \frac{\delta G_{nq}( t_2,t_2\,^+\, ) }{ \delta \phi_{km}(t_1)} \\
 \, &\,  \\
 \,             &\,\, =\,\, \sum_{ij} \frac{\delta G_{nq}( t_2,t_2\,^+ ) }{ \delta U_{ij}(\overline{t_3})}\times
                \frac{\delta U_{ij}(\overline{t_3}) }{ \delta \phi_{km}(t_1)} \\
\, &\,  \\
               &\,\, =\,\, \sum_{ijrs} G_{nr}(t_1,\overline{t_4})G_{sq}(\overline{t_4},\overline{t_5})
               \Gamma_{rs;ij}(\overline{t_4},\overline{t_5};t_2\,^+)\times
               \frac{\delta U_{ij}(\overline{t_3}) }{ \delta \phi_{km}(t_1)} \\
\end{array}\label{eq:pider}\eeq
\noindent where the definition of $\Gamma$ in Eqs. (\ref{eq:wgam}) and Eq. (\ref{eq:fdinv}) have been used. When
Eq. (\ref{eq:phon}) is substituted in Eq. (\ref{eq:pider}) one gets the first one of the constitutive Eqs.
(\ref{eq:pigam}). In Eq. (\ref{eq:pider}) $t_2\,^+ = t_2 + \eta$, where $\eta$ is a positive infinitesimal, to
ensure the correct order of the creation and annihilation operators in the equal times Green' s function. For the
vertex function $\Gamma$ one gets from Eq. (\ref{eq:Gm1})
\beq\begin{array}{ll}
\Gamma_{nh ; ij}(t_1,t_2 ; t_3) &\, =\, - \big( \frac{\delta G_{nh}^{-l}(t_1,t_2) }{ \delta U_{ij}(t_3)}\big) \\
\, &\,   \\
\, &\, =\, \delta_{ri}\delta_{hj}\delta(t_1 - t_3)\delta(t_2 - t_3) \, -\,
\big( \frac{\delta \tilde{M}_{nh}(t_1,t_2) }{ \delta U_{ij}(t_3)} \big)  \\
\, &\,   \\
\, &\, =\,  \delta_{ri}\delta_{hj}\delta(t_1 - t_3)\delta(t_2 - t_3) \, +\,
\sum_{rspq} \big( \frac{\delta \tilde{M}_{nh}(t_1,t_2) }{ \delta G_{rs}(\overline{t_4},\overline{t_5})} \big)
 G_{rp}(\overline{t_4},\overline{t_6})G_{qs}(\overline{t_7},\overline{t_5})
\big(\frac{\delta G_{pq}^{-l}(\overline{t_6},\overline{t_7}) }{ \delta U_{ij}(t_3)}\big) \\
\end{array}\label{eq:gamder}\eeq
\noindent and with the definition for $\Gamma$ in Eqs. (\ref{eq:wgam}) one obtains the second of the constitutive
Eqs. (\ref{eq:pigam}). This completes the derivation of the three basic equations of the formalism.

\vfill\eject

\vfill\eject

\vfill\eject

\vfill\eject

\vfill\eject
\vfill\eject

\vfill\eject

\end{document}